# On the meaning of an additional hypothesis in the Bell's inequalities.

Alejandro A. Hnilo.

CEILAP, Centro de Investigaciones en Láseres y Aplicaciones, UNIDEF(MINDEF-CONICET).
CITEDEF, J.B. de La Salle 4397, (1603) Villa Martelli, Argentina.
*emails*: ahnilo@citedef.gob.ar, alex.hnilo@gmail.com

*Abstract*

*The Bell's inequalities are derived from the hypotheses of Locality, Realism and (what is lesser known) the equality between the factual and the counterfactual time averages of the expectation values of observables. The necessity of a hypothesis additional to Local Realism opens a promising way out to the old controversy between Quantum Mechanics and Local Realism. For, it is possible to speculate that it is this additional hypothesis, and not Local Realism, what is disproved in the experiments reporting the violation of the Bell's inequalities. Yet, there are doubts on how the additional hypothesis may be violated in a physically reasonable process. A simple example showing that this is possible, the relationship between the validity of the additional hypothesis and the validity of the Bell's inequalities, and considerations on its physical meaning, are presented.*

February 24th, 2014.

**1. Introduction.**

The Bell's inequalities [1] are correlation bounds between the results of measurements performed on two particles. They are derived by consistently following a classical description of an experiment like the one in the Fig.1. This classical description holds to the innate ideas of Locality (roughly speaking: the results of an experiment are independent of the events outside its past lightcone) and Realism (r.s.: the properties of the physical world are independent of whether it is being observed or not). From these two innate ideas, two specific assumptions are derived: *measurement independence*, i.e., that the values of an hypothetical ("hidden") variable carried by the particles are uncorrelated from the setting of the observation apparatus (the $\alpha,\beta$ angles in Fig.1); and *counterfactual definiteness*, i.e., that it is possible to assign values to the results of non-performed observations. In the case of the setup in the Fig.1, the bound most used in the experiments is the Clauser-Horne-Shimony and Holt (CHSH) inequality, which reads:

$$S_{CHSH} \equiv |E(\alpha,\beta) - E(\alpha,\beta')| + |E(\alpha',\beta') + E(\alpha',\beta)| \leq 2 \qquad (1)$$

where the $E(\alpha,\beta)$ are expectation values defined as:

$$E(\alpha,\beta) = \{(+1).[C^{++} + C^{--}] + (-1)[C^{-+} + C^{+-}]\}/\{C^{++} + C^{--} + C^{-+} + C^{+-}\}|_{\alpha,\beta} \qquad (2)$$

where the $C^{ij}|_{\alpha,\beta}$ are the number of coincidences observed at the detectors if the analyzers are set to the values $\{\alpha,\beta\}$, and $i,j = + (-)$ means that a photon has been detected in the transmitted (reflected) output of the analyzer. The predictions of usual Quantum Mechanics (QM) apparently violate the CHSH inequality. F.ex., for the entangled state of two particles $|\varphi^+\rangle = (1/\sqrt{2})\{|x_a,x_b\rangle + |y_a,y_b\rangle\}$, if $\{\alpha,\beta,\alpha',\beta'\}$ are chosen equal to $\{0,\pi/8,\pi/4,3\pi/8\}$, then $S_{CHSH} = 2\sqrt{2} > 2$. This prediction (or others alike) has been experimentally confirmed under increasingly stringent conditions [2-5]. Therefore, leaving aside some remaining technical objections, the conclusion of most of the scientific community is that QM is in contradiction with Locality and/or Realism (in short: Local Realism, LR). This contradiction is a crucial issue, because LR is assumed not only in everyday life, but also in all the scientific practice (excepting QM).

Nevertheless, a detailed analysis shows that there is another hypothesis, which is necessary to insert the results of real measurements into the usual Bell's inequalities [6]. This additional hypothesis, which I call here "homogeneous dynamics assumption" (HDA) is not related with an experimental imperfection (as in the case of the "loopholes" [1]), but with the impossibility of measuring with two different angle settings at the same time. Taking into account the consequences of giving up LR, it seems sensible to speculate that the observed violation of the Bell's inequalities is a refutation of the HDA (or its related hypotheses, see later), not of LR. Yet, doubts have been raised on how any reasonable physical process may violate the HDA. In this paper, I present the example of a simple model that violates the HDA. It also illustrates the relationship between the violation of the HDA and that of the Bell's inequalities.

In the next Section, the reasoning leading to the necessity of the HDA is briefly reviewed for the CHSH inequality only. An extensive discussion can be found in [6]. The relationship of the HDA with the other possible hypotheses that lead to the usual form of the Bell's inequalities is commented. In the Section 3, the simple model violating the HDA is described. I warn the Reader that the aim here is *not* to propose a LR model able to violate the Bell's inequalities, but to study the physical meaning of the premises necessary for the validity of the usual way the experimental data are inserted into the Bell's inequalities.

In summary, the questions to be answered in this paper are:

1) What is the HDA? Why is it important?
2) Is there any physically reasonable situation where the HDA is violated?
3) What is the relationship between the violation of the HDA and the violation of the Bell's inequalities?

## 2. The homogeneous dynamics assumption (HDA).

*2.1 Review of the derivation of the CHSH inequality.*

The strength of the Bell's inequalities is that they take into account *all the possible ways* a classical system can produce correlated results between measurements performed on two remote particles. In order to produce such correlations, each particle is supposed to carry a "hidden variable", named $\lambda$, that influences the process of detection. In the case of the CHSH inequality, the observable corresponding to a coincident detection depends on the angle settings and $\lambda$, i.e.: $AB(\alpha,\beta,\lambda) \equiv A(\alpha,\lambda) \times B(\beta,\lambda)$ (where $A,B = \pm 1$) then:

$$AB(\alpha,\beta,\lambda) = \{(+1).[C^{++} + C^{--}] + (-1)[C^{-+} + C^{+-}]\}/\{C^{++} + C^{--} + C^{-+} + C^{+-}\}|_{\alpha,\beta,\lambda} \quad (3)$$

and its expectation value (i.e., the actually measurable value) is the average over the $\lambda$-space:

$$E(\alpha,\beta) = \int d\lambda . \rho(\lambda) . AB(\alpha,\beta,\lambda) \quad (4)$$

where $\int d\lambda . \rho(\lambda) = 1$, $\rho(\lambda) \geq 0$. For the angle settings $\{\alpha,\beta\}$ and $\{\alpha,\beta'\}$:

$$E(\alpha,\beta) - E(\alpha,\beta') = \int d\lambda.\rho(\lambda).AB(\alpha,\beta,\lambda) - \int d\lambda.\rho(\lambda).AB(\alpha,\beta',\lambda) = \qquad (5)$$
$$= \int d\lambda.\rho(\lambda).[AB(\alpha,\beta,\lambda) - AB(\alpha,\beta',\lambda)]$$

After adding and subtracting the term $[AB(\alpha,\beta,\lambda).AB(\alpha',\beta,\lambda).AB(\alpha,\beta',\lambda).AB(\alpha',\beta',\lambda)]$ inside the integral, reordering and applying the modulus (note that $1 \pm AB \geq 0$) [1]:

$$|E(\alpha,\beta) - E(\alpha,\beta')| \leq \int d\lambda.\rho(\lambda).[1 \pm AB(\alpha',\beta',\lambda)] + \int d\lambda.\rho(\lambda).[1 \pm AB(\alpha',\beta,\lambda)] = \qquad (6)$$
$$= 2 \pm [E(\alpha',\beta') + E(\alpha',\beta)]$$

that leads to the CHSH inequality, eq.1. Let see now why an additional hypothesis is necessary to apply this inequality to the data extracted from an experiment.

*2.2 The necessity of an additional hypothesis.*

All real measurements occur successively, during time. The expectation value is then:

$$E(\alpha,\beta) = (1/\Delta T)\int_{\theta}^{\theta+\Delta T} dt.\rho(t).AB(\alpha,\beta,t) \qquad (7)$$

which represents the following process: set the analyzers' angles to α and β during the time interval $[\theta,\theta+\Delta T]$, sum up the number of coincidences detected after each analyzer's outputs, and obtain $E(\alpha,\beta)$ from eq.3. If measurement independence is assumed valid, the distribution of the time intervals among the angle settings (i.e., if $[\theta,\theta+\Delta T]$ is a single continuous interval or many separated small intervals, see Fig.2) is irrelevant. Let take then, for simplicity, the distribution in the upper part of the Fig.2. A different distribution requires a more involved notation of the integration intervals, but the result is the same. Now, a problem becomes evident in the passage from the first to the second line of eq.6:

$$E(\alpha,\beta) - E(\alpha,\beta') = (1/\Delta T)\int_{T/4}^{T/2} dt.\rho(t).AB(\alpha,\beta,t) - (1/\Delta T)\int_{0}^{T/4} dt.\rho(t).AB(\alpha,\beta',t)$$
$$\neq (1/T)\int_{0}^{T} dt.\rho(t).[AB(\alpha,\beta,t) - AB(\alpha,\beta',t)] \qquad (8)$$

The rhs in the first line is what is actually measured, while the integral in the second line is the expression that leads to the CHSH inequality. The reason of the inequality is, simply, that the integration intervals are different. The same happens with the integrals for the inserted term [6]. In consequence, the measured values may violate the eq.1 or not but, at this point, this result implies nothing about LR. To restore the logical link between eq.1 and the LR hypotheses, it is necessary to sum all the observables under the *same* interval of integration (which is the λ-space in the usual derivation of the inequality). In order to do that, the integrals must be completed with expectation values obtained under conditions that did not occur, i.e., counterfactual values. This is not a problem, for Realism (more precisely: counterfactual definiteness), which is already assumed in the derivation, does allow counterfactual reasoning. The eq.8 becomes then:

$$E(\alpha,\beta) + \underline{E(\alpha,\beta)} - E(\alpha,\beta') - \underline{E(\alpha,\beta')} = (1/\Delta T)\int_0^T dt.\rho(t).[AB(\alpha,\beta,t) - AB(\alpha,\beta',t)] \qquad (9)$$

where the underlined terms are the sum of three counterfactual time averages:

$$\underline{E(\alpha,\beta)} = (1/\Delta T)\int_0^{T/4} dt.\rho(t)A\underline{B}(\alpha,\beta,t) + (1/\Delta T)\int_{T/2}^{3T/4} dt.\rho(t)A\underline{B}(\alpha,\beta,t) + (1/\Delta T)\int_{3T/4}^{T} dt.\rho(t)\underline{A}B(\alpha,\beta,t) \qquad (10)$$

$$\underline{E(\alpha,\beta')} = (1/\Delta T)\int_{T/4}^{T/2} dt.\rho(t)AB(\alpha,\beta',t) + (1/\Delta T)\int_{T/2}^{3T/4} dt.\rho(t)\underline{AB}(\alpha,\beta',t) + (1/\Delta T)\int_{3T/4}^{T} dt.\rho(t)\underline{A}B(\alpha,\beta',t) \qquad (11)$$

The (f.ex.) factor $A\underline{B}(\alpha,\beta,t)$ indicates the result of a measurement performed at a time value when B≠β, i.e., a counterfactual result (note that there are also double counterfactuals). By definition:

$$(1/\Delta T)\int_0^{T/4} dt.\rho(t).A\underline{B}(\alpha,\beta,t) = \{(+1)[C^{++}+C^{--}] + (-1)[C^{-+}+C^{+-}]\}/\{C^{++}+C^{--}+C^{-+}+C^{+-}\}\big|_{\alpha,\beta,0<t<T/4} \qquad (12)$$

which is unknown, because all the $C^{ij}$=0. Yet, counterfactual definiteness ensures that the rhs of eq.12 does take *some* value. Then the reasoning follows as usual, but the final expression is:

$$\left|E(\alpha,\beta) + \underline{E(\alpha,\beta)} - E(\alpha,\beta') - \underline{E(\alpha,\beta')}\right| + \left|E(\alpha',\beta') + \underline{E(\alpha',\beta')} + E(\alpha',\beta) + \underline{E(\alpha',\beta)}\right| \leq 8 \qquad (13)$$

I stress that it is eq.13, not eq.1, the inequality that can be derived assuming LR *only*. The eq.13 is not as trivial as it may appear, for the counterfactual terms span over three time intervals instead of one, and hence they can be, in principle, thrice larger than the corresponding factual ones.

As the Bell's inequalities are derived within LR, counterfactual reasoning is acceptable and no additional hypotheses are necessary to derive the eq.13. But, there is still the problem of assigning values to the counterfactual terms. In order to solve this problem, a "possible world" must be defined to ensure logical consistency [7]. There is no mystery in this situation, but simply lack of information. Let see an example of everyday life: let suppose that when I go to the pub and find my friend Alice there, the result of measuring the variable A ≡ "I find Alice in the pub" is 1, and 0 when I do not find her there. After many visits to the pub, I measure the expectation value ⟨A⟩=0.3. Now, let consider the question: "what is the expectation value of A when I don't go to the pub?" If it is assumed that Alice and the pub have a well defined existence even when I do not go there (roughly speaking, if Realism is assumed), the expectation value of A is some well defined number, say, *q*. But, the value of *q* cannot be known with the information available at this point. More information is needed, regarding the behavior of Alice and the properties of the pub when I am not observing them (i.e.: a "possible world" must be defined) to assign a numerical value to *q*.

Once a possible world is defined, the counterfactual terms can be calculated. Depending on the possible world chosen, the eq.1 is retrieved, or not [6]. The important point here is that the definition of a possible world unavoidably involves one assumption *additional* to LR. This weakens the consequences of the violation of the Bell's inequalities reported in the experiments, for the violation can be interpreted as a refutation of the additional hypothesis, not necessarily of LR. Note that this weakening does not arise from an experimental imperfection (as in the case of the loopholes [1]). The setup is assumed ideally perfect. The weakening arises from the fact that real measurements are performed during time, and that it is impossible to measure with two different angle settings at the same time. For, time is a magnitude that takes any given value only once.

It is conceivable that the weakening is the consequence of a weak point in the usual way the theory of probability is applied to the results of observations. In the theory, events are thought to occur in abstract, independent parallel worlds. The average over an ensemble of these parallel worlds allows the simple calculation of probabilities. In any actual observation instead, events occur (and averages are obtained) successively in time, in the only available real world. The equality between ensemble and time average values is known as the ergodic hypothesis [8]. The idea that non-ergodicity may be related with the solution of the QM vs LR controversy is not new [9-11].

*2.3 Conditions for retrieving the usual CHSH inequality, eq.1.*

The simplest way to retrieve eq.1 from eq.13 is to suppose a possible world where the factual and counterfactual observables are equal: $\underline{AB(\alpha,\beta,t)} = AB(\alpha,\beta,t)$ in eqs.10-11 (and the same for all the other counterfactuals). In the example of the pub: this possible world means that, *each day*, Alice would have been there, or not, regardless whether I went to the pub or not. This choice has, however, some technical drawbacks [6]. A less restrictive and, in my opinion, more appealing alternative, is to suppose that the factual and counterfactual *time averaged* expectation values are equal (in the example of the pub, this means that $q = \langle A \rangle = 0.3$). Each of the three counterfactual terms in the rhs in eqs.10 and 11 is then equal to the factual term, so that in eq.13:

$$\underline{E(\alpha,\beta)} = 3 \times E(\alpha,\beta) \tag{14}$$

the same for the other counterfactuals, and the eq.1 is retrieved. This alternative defines the HDA. Therefore, the correct logic inference is:

$$\text{LR} + \text{HDA} \Rightarrow S_{\text{CHSH}} \leq 2 \tag{15}$$

The ergodic hypothesis, mentioned before, means that the time averages are equal to the ensemble average. As the latter is unique, it implies that all the time averages (both factual and counterfactual) are equal among them. Therefore, the ergodic hypothesis implies the HDA. Yet, the HDA is different from the ergodic hypothesis, for it is conceivable that the factual and the counterfactual time averages are equal among them, but that they are different from the ensemble average. It is also conceivable that eq.14 holds even if the HDA is not valid. For, the three counterfactual integrals in eqs.10-11 may be all different, and still their sum may be equal to thrice the factual one. The different logical implications involved are summarized in the Fig.3. The sets of hypotheses painted in grey retrieve the usual Bell's inequalities. In the "other possible worlds" (i.e., outside the grey set) there is no logical link between the violation of the Bell's inequalities and LR.

As it is seen in the Fig.3, the HDA does not define the only possible world where the usual Bell's inequalities are retrieved. Yet, I believe that the HDA is the most plausible one in physical terms. It means that the time average of a dynamical variable, recorded during the interval $[t_1, t_1+T]$ (say, when that variable is being observed) is equal to the time average recorded during any other interval $[t_2, t_2+T]$ (say, when that variable is not being observed) provided, of course, that $T$ is sufficiently long. In fact, the HDA seems so plausible, that doubts were raised on the existence of a physical process able to violate it. In other words: there are no doubts that possible worlds can be defined that violate the HDA (some of them are discussed in [6]). The question is whether some reasonable mechanism exists that, as a *result* of its evolution, violates the HDA. It is desirable that such mechanism also violates the Bell's inequalities, and that it does so for some (not all) the values of its parameters, in order to be useful to study the relationship between the validity of the HDA and that of the Bell's inequalities. In the next Section, such mechanism is presented.

## 3. A simple model that violates the HDA.

*3.1 A classical mixture and delay.*

The QM predictions for the state $|\varphi^+\rangle$ can be reproduced, regardless the position of the analyzer in station B, by a statistical mixture of photon pairs polarized parallel and orthogonal to the analyzer in station A, or:

$$\rho_\alpha = \tfrac{1}{2} \{|\alpha\rangle\langle\alpha|_A \otimes |\alpha\rangle\langle\alpha|_B + |\alpha_\perp\rangle\langle\alpha_\perp|_A \otimes |\alpha_\perp\rangle\langle\alpha_\perp|_B\} \tag{16}$$

This state is classical, it is able to reproduce the QM predictions only because the value of $\alpha$ is known by the source. In the general case that $\{a,b\}$ are the setting angles of the analyzers and $\alpha$ can be equal to $a$ or not, the probability of coincidence produced by $\rho_\alpha$ is:

$$P^{++}(a,b,\alpha) = \tfrac{1}{2} \{\cos^2(a-\alpha)\cdot\cos^2(b-\alpha) + \sin^2(a-\alpha)\cdot\sin^2(b-\alpha)\} \tag{17}$$

which is the probability for the pair passing the analyzers as $x$-polarized or as $y$-polarized photons. As $P^{++}+P^{+-}=\tfrac{1}{2}$ and $P^{++}=P^{--}$, then $E(a,b,\alpha) = 4\times P^{++}(a,b,\alpha)-1$ if the rate of coincidences is arbitrarily high. Assume now that $\alpha$ evolves in time according to the following equation:

$$d\alpha(t)/dt = -(\Gamma/\tau)[\alpha(t-\tau) - a(t)] \tag{18}$$

where the function $a(t)$ is externally defined. F.ex., it is the randomly varying angle setting in [3]. The argument of $\alpha$ in the rhs is delayed a time $\tau \geq L/c$ to take into account the time required by the information (on the value of the angle setting) to cover the spatial spread of the setup. The eq.18 holds to Locality, but not necessarily to measurement independence. The largest meaningful distance between $\alpha$ and $a$ is $\pi/4$, so that $\alpha$ is scaled to the $[-\pi/4,\pi/2]$ interval. The value of $\alpha$ is always well defined, what allows the calculation of both the factual and counterfactual $E(a,b)$. The parameter $\Gamma$, whose value is unknown, measures the strength of the tracking force that makes $\alpha(t)$ to follow the values taken by $a(t)$. The QM predictions are fully retrieved in the limit $\tau \to 0$.

The eq.18 is a delay differential equation. It evolves in a phase space of infinite dimensions. In the particular case that $a(t)$ = constant $\neq \alpha(t=0)$, $\alpha(t)$ decays slowly to $a$ if $\Gamma \ll 1$. If $\Gamma \approx 1$, $\alpha(t)$ decays to $a$ fast, with damped oscillations of period $\approx 4\tau$. If $\Gamma \geq \pi/2$, the amplitude of the oscillations of $\alpha(t)$ increases exponentially [12]. Be warned that the model proposed in [12] is similar, but different, from the one presented here.

*3.2 The static case.*

Let consider first the distribution of the measuring time sketched at the top of Fig.2, which is the case in most of the performed experiments. If $T>>\tau$ and $\Gamma$ is not too small, then $\alpha(t)=a(t)$ during practically all the time, the QM predictions are reproduced and $S_{CHSH}\approx2\sqrt{2}$. From eq.17, one of the three counterfactual terms (the one with the actual value of *a*) in the rhs of eqs.10-11 is equal to the factual time average, and the other two terms are zero. Hence, $\underline{E(a,b)} = E(a,b) = \pm\sqrt{2}/2$. The eq.14 is thus not fulfilled, and the usual CHSH inequality is not retrieved. The HDA is violated. In consequence, if the values of the $E(a,b)$ are inserted into the eq.1, $S_{CHSH}>2$ (as it is said before), but this result implies nothing about the validity of LR. It is convenient defining a variable $\Delta$ to quantify the violation of the HDA:

$$\Delta \equiv \sum_{i,j} \frac{|E(i,j) - (1/3)\underline{E(i,j)}|}{|E(i,j)| + (1/3)|\underline{E(i,j)}|} \qquad (19)$$

where *i,j* are the four usual setting angles. If the HDA holds, $\Delta=0$. In this static case, $\Delta=2$. The usual CHSH inequality is retrieved iff $\Delta=0$.

*3.3 The random variable case.*

The case when the angle settings are varied randomly (the lowest one in the Fig.2) is illustrative. A numerical simulation of eq.18 is carried out, with $a(t)$ jumping randomly, with transition time equal to zero, between the values 0 and $\pi/4$. The jump times are chosen also randomly, at an average rate $\mu$ [13]. The factual and the counterfactual time averages are computed. Each value of $S_{CHSH}$ (from eq.1) and $\Delta$ is obtained after $2000\tau$ with a coincidence rate of $500/\tau$ (total coincidences=$10^6$) and after discarding a transient of $200\tau$ (initial condition $a(t<0)=0$). The rate of incident pairs is high enough to follow the variations of $\alpha(t)$ in detail; this condition is far from having been reached in any experiment. Due to the fluctuations, the strict equality $\Delta=0$ is never obtained numerically. In what follows, I will say that the HDA holds if $\Delta<<1$ ($\Delta\approx0$), and that it is violated if $\Delta\approx1$. For some parameters' values, $S_{CHSH}>2$ is obtained. In these cases, it is always found that $\Delta\approx1$, i.e, the HDA is violated too. The inverse is not true: it is possible that $\Delta\approx1$ and yet $S_{CHSH}<2$. In other words: the violation of the HDA is a condition necessary, but not sufficient, for the violation of the usual form of the Bell's inequalities.

F.ex., in the Fig.4 the variation of $S_{CHSH}$ and $\Delta$ as a function of $\Gamma$ are shown for $\mu\tau=$ ¼. The vertical dotted lines at $\Gamma\approx0.2$ and $\Gamma\approx\pi/2$ are the limits of the region where the CHSH inequality is violated. In this region, $\Delta\approx1$. Besides, the shape of the curve of $\Delta$ follows the one of $S_{CHSH}$. For $\Gamma<0.2$ one gets $S_{CHSH}<2$, but it is still $\Delta\approx1$. For $\Gamma\to0$, $\Delta\to0$ and $S_{CHSH}\to\sqrt{2}$, which is the value produced by $\rho_\alpha$ if there is no correlation between $\alpha(t)$ and $a(t)$. At $\Gamma\approx\pi/2$ instead, $S_{CHSH}$ falls to $\sqrt{2}$ and $\Delta$ falls to zero abruptly. The reasons of these numerical results are understood from the inspection of $\alpha(t)$. For $\Gamma=1$ (Fig.5a) $\alpha(t)$ promptly adjusts to the changing values of $a(t)$, the condition $\alpha(t)\approx a(t)$ holds most of the time and hence $S_{CHSH}>2$. For $\Gamma=0.1$ (Fig.5b) $\alpha(t)$ follows $a(t)$ too slowly. It reaches the target values 0 and $\pi/4$ only rarely and, in consequence, $S_{CHSH}<2$. The fact that $\Delta\approx1$ in these two cases is a non trivial result. For $\Gamma=0.02$ (Fig.5c) the tracking force is too weak, and $\alpha(t)$ just makes a low amplitude zigzag around the middle value $\pi/8$. Therefore, $\alpha(t)$ and $a(t)$ are mostly uncorrelated and $S_{CHSH}\approx\sqrt{2}$. As $\alpha(t)$ is nearly constant, the HDA is valid and thus $\Delta\approx0$. At $\Gamma=2$ (Fig.5d) $\alpha(t)$ diverges exponentially, rotating faster and faster. The jumps of $a(t)$ make the evolution even wilder, reversing the runaway at random times. In consequence, $\alpha(t)$ loses any correlation with $a(t)$. This explains why $S_{CHSH}$ falls abruptly to $\sqrt{2}$ at $\Gamma\approx\pi/2$. The trajectory of $\alpha(t)$ randomly explore its whole phase space, apparently fulfilling the condition of "mixing" that makes the ergodic hypothesis valid [8]. Thus, it is not surprising that $\Delta$ falls to zero abruptly at this point.

This simple model does not violate the CHSH inequality (nor the HDA) if $\mu\tau\gg1$. There are several ways to achieve that [14-17]. But, in my opinion, these complications would obscure the ideas of interest here. As it is said, the aim here is not to get a LR model that violates the Bell's inequalities, but to explore the meaning and consequences of the HDA and its related hypotheses.

**4. Summary and discussions.**

The questions made at the end of the Introduction can be answered now:

*1) What is the HDA? Why is it important?*

The usual Bell's inequalities cannot be applied to measurements that occur during time (i.e., all real measurements, even in an ideally perfect setup) unless at least one additional assumption is made. This assumption is the definition of a "possible world" to calculate the numerical values of the counterfactual terms in the inequality that can be deduced assuming LR only. The equality between the time averages of factual and counterfactual expectation values defines the HDA, which is one of the possible worlds that allow retrieving the usual Bell's inequalities. Yet, there is no fundamental reason why the HDA must be valid in the setup in Fig.1. It is reasonable then to speculate that is the HDA, and not LR, what is disproved in the experiments reporting the violation of the Bell's inequalities.

*2) Is there any physically reasonable situation where the HDA is violated?*

Yes. Assume that the source in the Fig.1 emits a mixture of photons polarized parallel and perpendicular to some angle $\alpha(t)$. Let $\alpha(t)$ try to follow the random variations of the angle set in the station A, after a delay $L/c$ and with tracking force of strength $\Gamma$. This example holds to LR and, for some values of the parameters, it violates the HDA.

*3) What is the relationship between the violation of the HDA and the violation of the Bell's inequalities?*

The presented example shows that the violation of the HDA is a condition necessary, but not sufficient, to violate the Bell's inequalities. This is a non trivial result.

The simple model presented here violates the HDA. Yet, if the HDA were actually violated in the experiments, the real physical process would be presumably different. The interesting question at this point is: how to know from the experimental data if the HDA is violated, even if the underlying physical process is unknown? Recall that the HDA is defined by the equality between factual averages (which are measured) and counterfactual ones (which cannot be measured).

If the simple model presented here is, at some extent at least, representative of the general case, a clue may be found in the time behavior. Note that a "too simple" $\alpha(t)$ (as if $\Gamma \ll 1$, Fig.5c), and a "too complex" one (as if $\Gamma \geq \pi/2$, Fig.5d), both hold to the HDA. It seems that the HDA is violated if the system has a "complex, but not too complex", behavior. Therefore, the complexity of a time series may be a way to detect if the HDA is violated. This way is far from being easy: a practically computable definition of complexity is a controversial and difficult issue. Besides, the time series in the Fig.5, that are taken here as a guide, assume a rate of pairs several orders of magnitude larger than what has been experimentally obtained. Last but not least, the hypothetical hidden variables, which would play the role of $\alpha(t)$ here, are unknown and not directly accessible. In spite of these difficulties, I believe that time resolved measurements (as opposed to the standard measurement of time averages only) are a promising way to detect a violation of the HDA. For, the embedding theorem ensures that an underlying attractor can be reconstructed from the scalar time series of any dynamical variable. The key problem is to find the appropriate variable, taking into account the practical limitations due to the precision of the measurement, the finite size of the series, the achievable sampling density, the noise, etc.

If the violation of the HDA in the setup of Fig.1 were experimentally confirmed, the arising solution of the QM vs LR controversy could look like this: QM would be the steady state approximation (i.e., the HDA holds) of a still unknown LR theory. This solution would be similar to the old "statistical interpretation" of QM, but augmented with non-ergodic evolution and dynamical complexity as essential ingredients.

**Acknowledgements.**

This work received support from the contract PIP CONICET 2011-01-077.

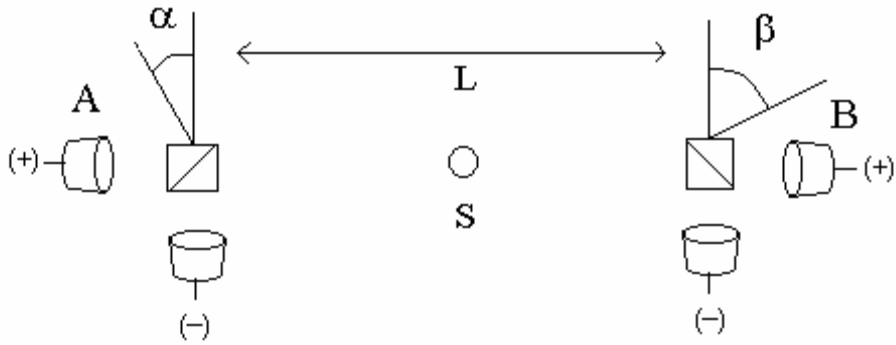

Figure 1: Scheme of a typical experiment to test the Bell's inequalities. The source S emits pairs of photons entangled in polarization towards stations A and B separated by a large distance *L*. At each station, the analyzer is set at some angle value. The expectation value *E($\alpha,\beta$)* is measured from the number of coincidences at each pair of outputs, recorded during some time interval. A photon detected in the transmitted (reflected) output of the analyzers is defined as a + (-) result.

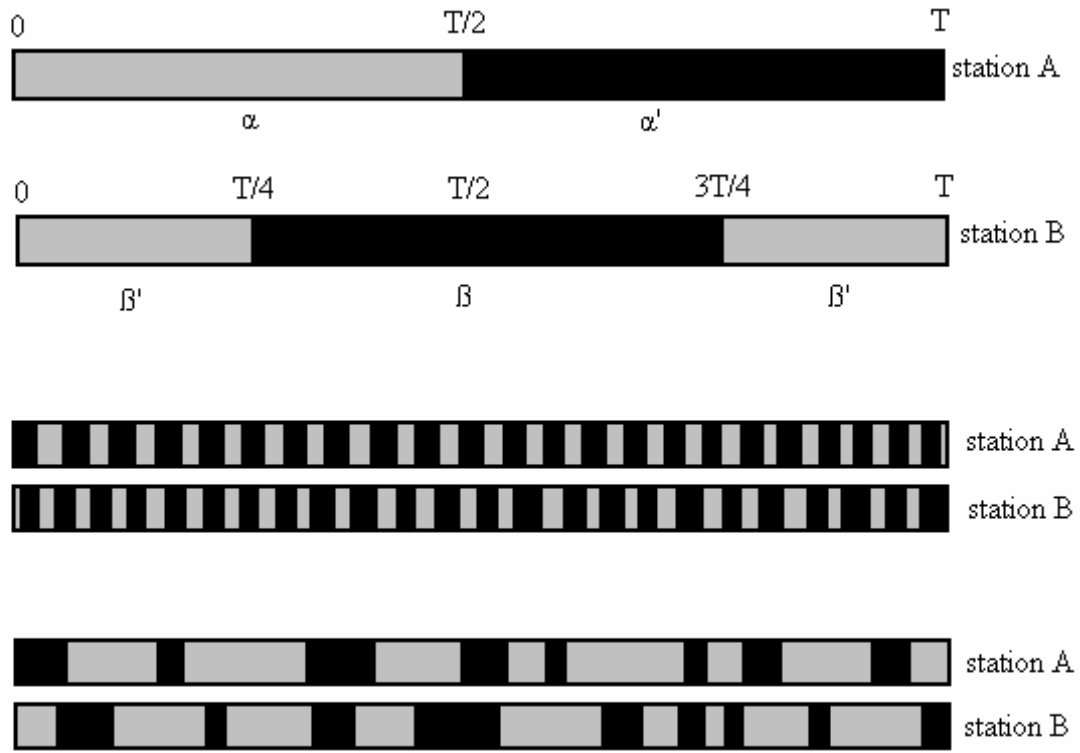

Figure 2: Schemes of three possible distributions of the measuring time among the different angle settings. Time intervals in grey: the angles are $\alpha$ and $\beta$'; in black: $\alpha$' and $\beta$. The flux of pairs is assumed constant in time. The upper distribution is the most usual in the experiments. It is, for simplicity, the one used in the text (measurement independence is assumed valid). The distribution in the center varies in a quasi-periodic way (as in [2]). The lowest one varies randomly (as in [3]). It is possible to translate one distribution into the other by merely changing the definition of the integration intervals.

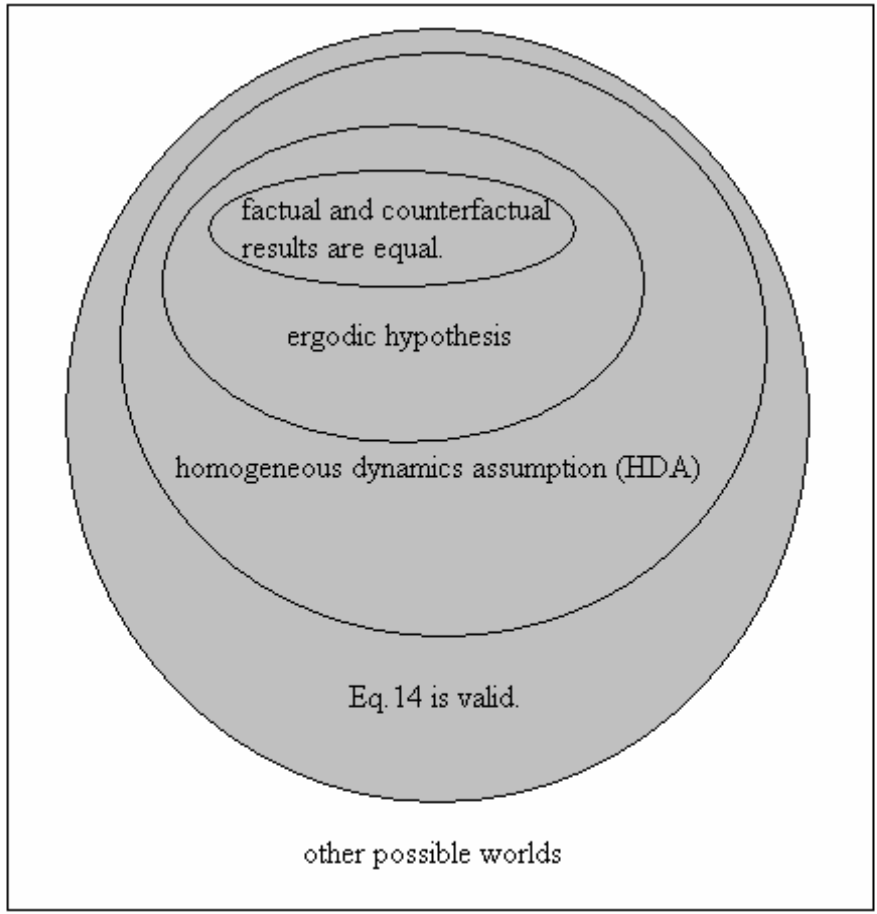

Figure 3: Logical relationship between different additional hypothesis defining different "possible worlds". F.ex.: in a possible world where the ergodic hypothesis is valid, the HDA and the eq.14 are also valid. In this same possible world, the equality between factual and counterfactual results may be valid, or not. The hypotheses in grey allow retrieving the usual Bell's inequalities. In the "other possible worlds" (some of them are discussed in [6]) there is no link between the violation of the usual Bell's inequalities and the validity of LR.

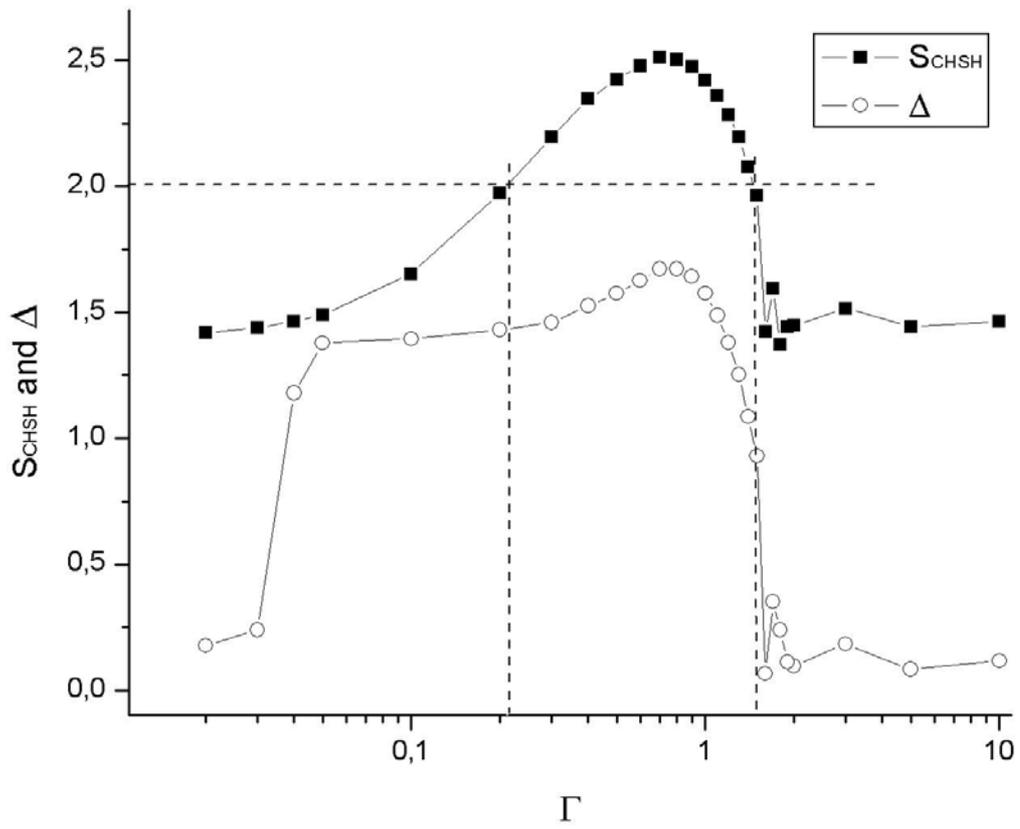

Figure 4: $S_{CHSH}$ and $\Delta$ as a function of $\Gamma$, $\mu\tau = ¼$. The horizontal dotted line indicates the CHSH bound.

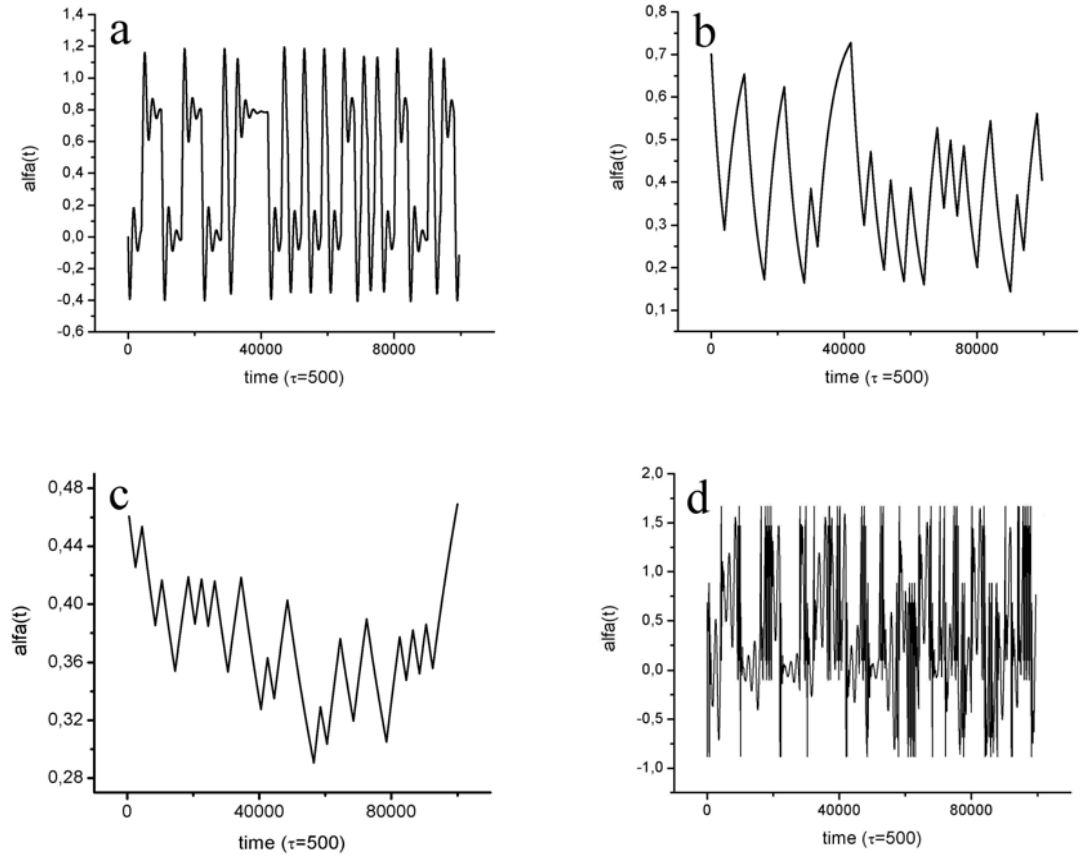

Figure 5: $\alpha(t)$ for $\mu\tau = ¼$ and several values of $\Gamma$, note the different vertical scales; (a) for $\Gamma=1$, $\alpha(t)$ promptly follows the random jumps of $a(t)$ between 0 and $\pi/4$, $S_{CHSH} > 2$ and $\Delta \approx 1$; (b) for $\Gamma=0.1$, $\alpha(t)$ follows $a(t)$ slowly, rarely reaching 0 or $\pi/4$, $\Delta \approx 1$ but $S_{CHSH} < 2$; (c) for $\Gamma=0.02$, $\alpha(t)$ zigzags around $\pi/8$ never reaching 0 or $\pi/4$, $S_{CHSH} \approx \sqrt{2}$ and $\Delta \approx 0$; (d) for $\Gamma=2$, $\alpha(t)$ varies wildly, filling the whole phase space and losing correlation with $a(t)$, $S_{CHSH} \approx \sqrt{2}$ and $\Delta \approx 0$.